**Investigation of phonons and magnons in [Ni$_{80}$Fe$_{20}$/Au/Co/Au]$_N$ multilayers**

M. Zdunek[1], S. Shekhar[1], S. Mielcarek[1], A. Trzaskowska[1*]

[1] Faculty of Physics, Adam Mickiewicz University in Poznań, Poznań, Poland

* E-mail: olatrzas@amu.edu.pl

**Abstract**

The interaction between phonons and magnons is a widely developing topic, especially in the field of acoustic spintronics. To discuss this interaction, it is necessary to observe two different waves (acoustic and spin waves) with the same frequency and wavelength. In the Ni$_{80}$Fe$_{20}$/Au/Co/Au system deposited on a silicon substrate, we observe the interaction between spin waves and surface acoustic waves using Brillouin light scattering spectroscopy. As a result, we can selectively control (activate or deactivate) the magnetoelastic interaction between the fundamental spin wave mode and surface acoustic waves by adjusting the magnetostrictive layer thickness in the multilayer. We demonstrate that by adjusting the number of layers in a multilayer structure, we can precisely control the dispersion of surface acoustic waves, with minimal impact on the fundamental spin wave mode.

1. **Introduction**

In conventional electronic devices, charge current is used for transmitting and processing information. To meet the increasing demand for smaller, more intelligent, and energy-efficient microwave devices, an alternative technology is urgently needed that goes beyond the current complementary metal-oxide-semiconductor (CMOS) technology based on charge current. Magnetic nanostructures have promising applications in magnonics and are a dynamically developing potential source of devices based on the use of spin waves for transmitting and processing information [1], [2], [3]. However, it should be emphasized that the characteristics of spin waves can be altered by a wide range of parameters, including the choice of magnetic material, sample shape, and the orientation and magnitude of the applied magnetic field [4]. It is important to note that in such devices, in addition to spin waves (magnons), acoustic waves (phonons) can also propagate [5], [6], [7], [8]. To design practical magnonic-phononic devices, it is necessary to control the frequency band structures in the wave vector domain for both spin and acoustic waves [9], [10]. Simultaneously, the examination of elastic properties and, therefore, the propagation of elastic waves in magnonic devices increases the functionality of these devices [11]. It is particularly interesting to consider the impact of elastic waves on spin waves. This can be achieved by tuning the aforementioned waves in the frequency and wave vector domain, creating conditions that lead to the effect of phonon-magnon interaction. Fine-tuning is possible through the appropriate selection of both magnetic parameters and elastic properties of the considered nanostructures.

A ferromagnetic layer deposited on an elastic and non-magnetic substrate enables the simultaneous propagation of surface acoustic waves (SAWs) and spin waves (SWs). This basic magnon-phonon hybrid system has a simple structure. However, dispersion branches of SAWs are observed at frequencies much lower than the ferromagnetic resonance (FMR) frequencies of the magnetic layer.

For the fundamental SW mode and the lowest standing SW modes perpendicular to the layers, the multilayer is treated as a homogeneous material with a reduced effective magnetic saturation value. To overcome this problem, a multilayer approach can be used, where thin magnetic and non-magnetic layers are arranged alternately. This reduction in magnetization lowers the frequencies of the magnon dispersion branches, allowing interaction with specific types of SAWs. The coupling between SW and SAW is weak [12], [13], [14]. Therefore, the (anti)crossings of magnon and phonon dispersion branches should be analyzed carefully. Additionally, the magnetoelastic interaction depends on the orientation of the wave vector relative to the direction of the static magnetic field, as well as the spatial profiles of the SAWs [15], [16].

The presented work analyses the magnon and phonon properties of multilayers with different repetitions ($N$) of $[Ni_{80}Fe_{20}/Au/Co/Au]_N$ deposited on a silicon substrate. The main objective of this study is to investigate the influence of the number of repetitions in a multilayer $[Ni_{80}Fe_{20}/Au/Co/Au]$ on the dispersion of SAWs, and subsequently on the magnetoelastic interaction with SW. The magnon and phonon dispersion in this system were determined and analyzed using Brillouin light scattering measurements and finite element method (FEM) calculations.

## 2. Materials and methods

### 2.1. The samples

The multilayer under investigation, denoted as $[Ni_{80}Fe_{20}/Au/Co/Au]_N$, consists of a specific number of repetitions ($N$) of permalloy layers (with a thickness of $t_{NiFe} = 2$ nm) and cobalt layers (with a thickness of $t_{Co} = 0.8$ nm), separated by gold layers (with a thickness of $t_{Au} = 2$ nm). These layers are deposited onto a gold buffer layer (with a thickness of $t_{BAu} = 30$ nm). The substrate used for the sample is naturally oxidized (100) silicon. The deposition process took place in an argon atmosphere using high vacuum magnetron sputtering on the substrate [17]. A schematic representation of the sample is depicted in Fig. 1a.

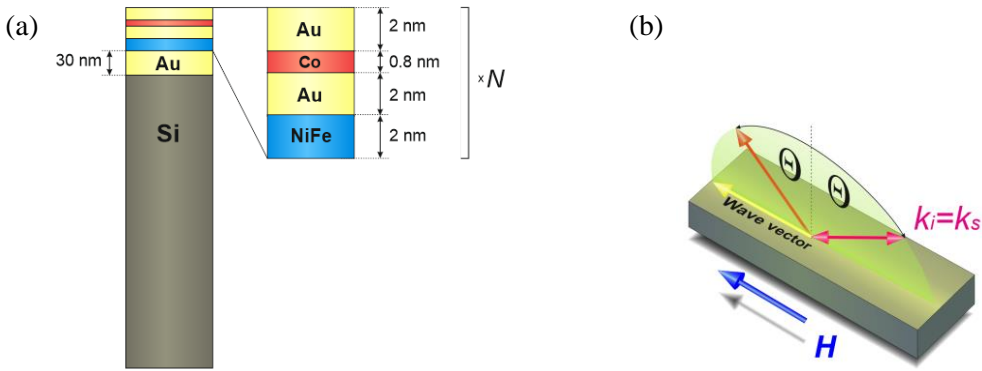

Fig. 1. (a) Schematic sketch of the considered system, composed of thick silicon (100) substrate with the gold buffer layer (30 nm) and magnetic multilayer. (b) Schematic representation of backward volume (BV) geometry. The magnetic field $H$ (blue arrow) is applied in the plane of the sample and is parallel to the plane of incidence for the light (photons) which interacts with magnons and phonons (pink arrow denotes the wave vectors $k_i$, $k_s$ for incidence and scattered light). The wavevector for SAWs

and SWs (yellow arrow) has the same direction as the applied field (backward volume (BV) geometry for SWs in $Ni_{80}Fe_{20}$).

*2.2. Experimental setup*

To obtain the dispersion relations of SAWs and SWs, we used a six-pass, tandem Brillouin spectrometer (TFP2-HC, JRS) which ensures a contrast of $10^{15}$ [22]. The source of scattering light was a frequency-stabilized DPSS laser, which operates at $\lambda_0 = 532$ nm (Coherent Verdi V5). The geometry used for those measurements was a 180° backscattering geometry with *pp* polarization for SAWs and *ps* polarization for SWs. The frequency of SAWs and SWs is represented by the Brillouin frequency shift of the inelastically scattered laser beam. Changing the angle of light incidence, $\Theta$, concerning the surface of the sample, allows selecting the wave vector $q = \frac{4 \cdot \pi \cdot \sin(\Theta)}{\lambda_0}$, common for SAWs and SWs, and determining the dispersion relation [18], [19], [20]. In our experiments, the wave vector was varied in the range 0.007-0.022 nm$^{-1}$ with a resolution of about 0.001 nm$^{-1}$. The free spectral range (FSR) was 20 GHz with a frequency resolution of about 0.04 GHz. The sample was placed in the external magnetic field (50 mT), was applied in the plane of the sample, which corresponds to backward volume (BV) geometry for SWs in permalloy (Fig. 1b). Each spectrum was accumulated for 6000 cycles. A Lorentzian curve was fitted for each peak. A more detailed description of the experimental setup can be found in [20].

*2.3. FEM simulations*

The dispersion relation for SAWs propagating in the studied sample was simulated using the finite element method based COMSOL Multiphysics with Acoustic module [21]. The multilayer structure had been treated as one uniform material with elastic constant were: $c_{11}=c_{22}=c_{33}=$ 213.6·10$^9$ N/m$^2$, $c_{44}=c_{55}=c_{66}=$ 44.7·10$^9$ N/m$^2$, $c_{12}=c_{13}=c_{23}=$ 124.1·10$^9$ N/m$^2$ and mass density $\rho = 14948$ kg/m$^3$. The main components of the elasticity tensor for buffer gold layer was: $c_{11}=c_{22}=c_{33}=$ 192.3·10$^9$ N/m$^2$, $c_{44}=c_{55}=c_{66}=$ 41.9·10$^9$ N/m$^2$, $c_{12}=c_{13}=c_{23}=$ 163.1·10$^9$ N/m$^2$, and density $\rho = 19300$ kg/m$^3$ while for silicon substrate it was: $c_{11}=c_{22}=c_{33}=$ 166.6·10$^9$ N/m$^2$, $c_{44}=c_{55}=c_{66}=$ 79.6·10$^9$ N/m$^2$, $c_{12}=c_{13}=c_{23}=$ 64·10$^9$ N/m$^2$ and mass density $\rho = 2331$ kg/m$^3$ [22], [23], [24]. The elastic constants and the mass densities of each material consider the crystallographic orientation. The idea of calculating the dispersion relation for phononic samples was presented in Ref. [12], [24], [25].

**4. Results and Discusion**

At the surface of the substrate with the multilayer deposited on it, the surface acoustic waves (SAWs) can propagate. The spin waves (SWs) of long wavelength can propagate in a magnetic multilayer structure. In multilayered systems under investigation, it is possible to examine both types of waves. The spectra obtained for the samples with repetition *N*=3 and 12 of a [$Ni_{80}Fe_{20}$/Au/Co/Au]$_N$ multilayer are presented in Fig. 2. The larger number of repetitions *N* of the multilayer [$Ni_{80}Fe_{20}$/Au/Co/Au]$_N$, results in an increase in the total thickness of the multilayer, which significantly affects the spectrum of SAWs.

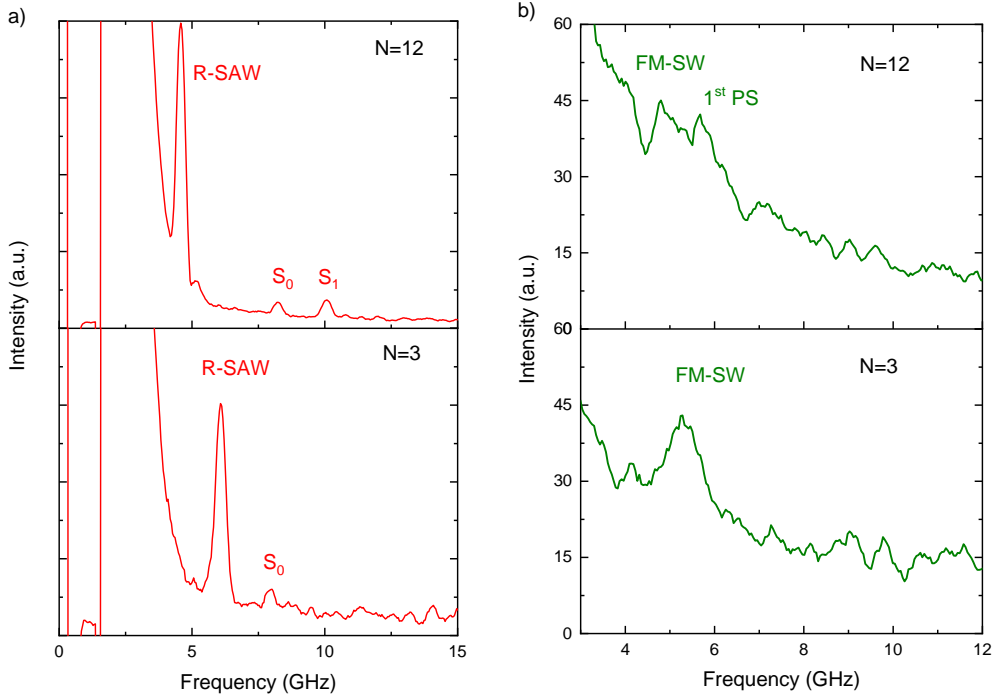

Fig. 2. (a) The BLS spectra for phonons depending on the number of repetitions $[Ni_{80}Fe_{20}/Au/Co/Au]_N$ for the wave vector $q= 0.01692$ nm$^{-1}$ with visible Rayleigh SAW (R-SAW) and Sezawa SAW ($S_0$, $S_1$). (b) Spin wave BLS spectra for $[Ni_{80}Fe_{20}/Au/Co/Au]_N$ multilayers differing in the number of repetitions $N$. The spectra were measured for the geometry supporting BV with a magnetic field 50 mT and wave vector 0.01360 nm$^{-1}$ with visible fundamental mode (FM-SW) and first perpendicularly standing mode (1$^{st}$ PS) for samples with different number of repetitions ($N=12$ and $N=3$).

The SAW with the lowest frequency corresponds to the Rayleigh surface acoustic wave (R-SAW), a type of wave commonly observed phenomena in all solid-state materials. Other modes, which are visible in the BLS spectra, are the Sezawa surface acoustic modes ($S_0$-$S_1$). The number of Sezawa SAWs observed within a fixed frequency range increases with thicker multilayers (i.e., multilayers with a higher number of repetitions, N). Sezawa SAWs are characteristic of the overlayer-substrate system [22], [26]. By considering the elastic properties of both the substrate and the complex overlayer (comprising buffer layers and multilayers), the studied structure can be classified as a "*slow-on-fast*" system [22], [27]. This characteristic can be confirmed by analyzing the dependence of the phase velocity of SAWs ($v$) on the wave vector ($q$) visible in Fig. 3b.

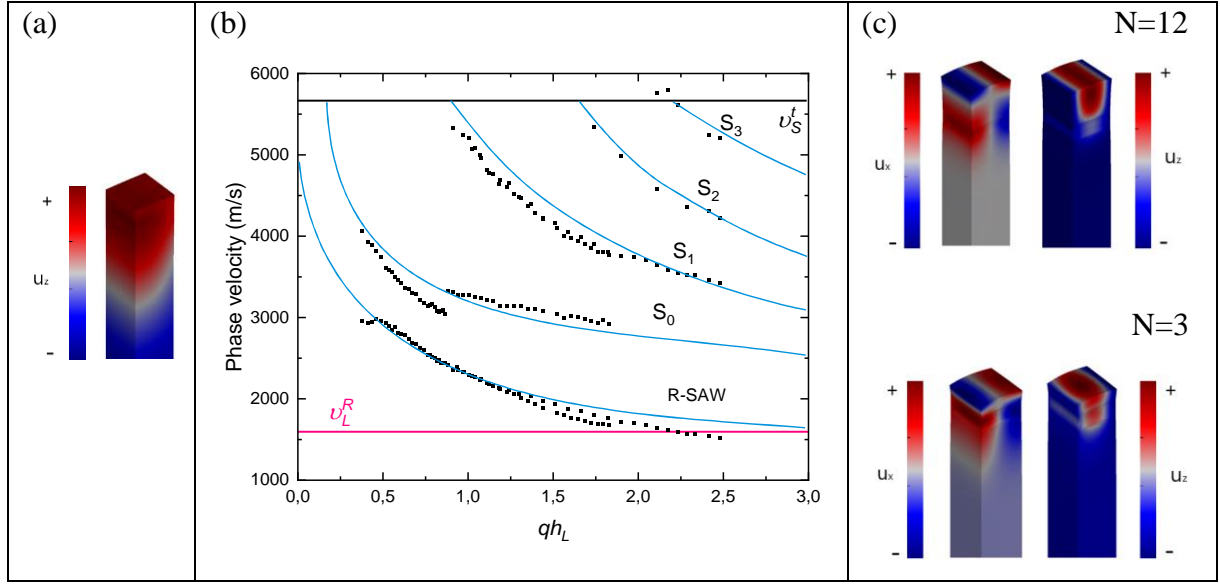

Fig. 3. The absolute value of the in-plane ($u_x$) and out-of-the plane ($u_z$) displacements for R-SAW for (a) $qh_L \to 0$ and (c) $qh_L \to 2$. (b) The phase velocity $v$ of SAWs in dependence on dimensionless wave number $qh_L$, where $h_L = N(t_{Au} + t_{NiFe} + t_{Au} + t_{Co}) + t_{BAu}$ stands for the thickness of the complex overlayer. The points and lines correspond to experimental and numerical outcomes for R-SAW and Sezawa SAWs ($S_0$-$S_3$). Please note that the experimental data were aggregated from all investigated samples, differing in the number of repetitions $N$. The horizontal lines $v_L^R$ $v_S^t$ mark the phase velocities of R-SAW and transverse bulk wave for overlayer (isolated from the substrate) and substrate (unloaded), respectively.

For large values of $qh_L$, the phase velocity approaches the value for isolated overlayer $v_L^R$. On the other hand, in the limit $qh_L=0$ the occupation of the overlayer by SAW is negligible compared to the substrate and the limiting phase velocity corresponds to the R-SAW velocity of bare substrate $v_S^R$. The negative slopes for all SAWs, visible in Fig. 3b, are consistent with the conditions $v_S^t > v_L^t$ for the *slow-on-fast* where $v_S^t$ and $v_L^t$ are velocities of transversal bulk waves in the substrate or in the layer, respectively. The numerical results, presented in Fig. 3b. were obtained by FEM calculations described in Ref. [22] whereas the experimental outcomes originate from the BLS measurements (phase velocity is a ratio of frequency and wave vector). Moreover it is possible to conclude that with the increase of the $qh_L$, the R-SAW becomes more localized. Information about localisatin of mode displacement can be inferred from simulations of the displacement components of the waves. The absolute value of the out-of-the plane and in-plane displacements for the R-SAW for different numbers of repetitions show strong localized modes in the multilayer system which is visible in Fig. 3a, c. This localization is larger for higher values of $qh_L$ while for small values of $qh_L$ R-SAW also penetrates into the substrate.

The magnetic characteristics of [$Ni_{80}Fe_{20}$/Au/Co/Au]$_N$ multilayers are notably intriguing. Unlike the in-plane magnetization of the permalloy layer, the cobalt layer within such structures exhibits out-of-plane magnetization. Consequently, the lowest-frequency SW eigenmodes are expected to be primarily localized within the $Ni_{80}Fe_{20}$ layer, characterized by its lower saturation magnetization and ferromagnetic resonance frequency. The cobalt subsystem is anticipated to act as a source of static stray fields, akin to the presence of perpendicular anisotropy in an in-plane magnetized permalloy layer. For a thinner sample, we can only see the fundamental mode of almost the same frequency and the perpendicularly standing (FM-SW) and first perpendicularly standing mode (1st PS). For a thinner sample, we can see the only fundamental mode of almost the same frequency – perpendicularly standing

SW modes were pushed up in frequency scale and thus, are not detected in the Brillouin spectrum (Fig. 2b).

The obtained dispersion relations of SAWs and SWs (for the configurations supporting BV modes) have been shown in Fig. 4. We showed the dispersion relation in a selected range of the wave vector to observe the area of anticrossing of the phononic and SW dispersion curves. Fig. 4 presents the results for the structures differing in the number of repetitions within the $[Ni_{80}Fe_{20}/Au/Co/Au]_N$ multilayers: $N=3$ in Fig. 4a and $N=12$ in Fig. 4b.

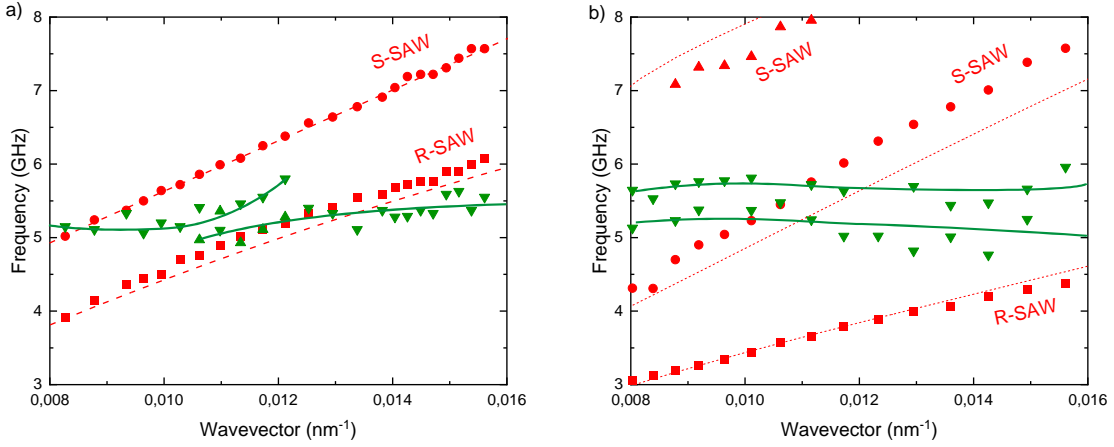

Fig. 4. Dispersion relations measured for SAWs (red points) and SWs (green points) for the BV geometry in structure with $N=3$ (a) and $N=12$ (b) repetitions of the $[Ni_{80}Fe_{20}/Au/Co/Au]_N$ multilayer. The red dashed line shows the calculated dispersion relations for R-SAW and the lowest Sezawa SAW (S-SAW). (a) The continuous green line is drawn, as a guide for the eye, to show the expected dispersion of hybridized (anticrossed) SW fundamental mode (FM-SW) and Love SAW (see Ref. [12]). (b) The continuous green line is drawn, as a guide for the eye, to show the expected dispersion fundamental SW (FM-SW) mode and the first perpendicular standing SW (1st PS) mode.

The increase of the thickness of the overlayer (due to the larger number of repetition $N$ in the $[Ni_{80}Fe_{20}/Au/Co/Au]_N$ multilayer), results in the a significant decrease of the phase velocity (and group velocity) for SAWs which is observed as a reduction of the slope of SAW dispersion branches in Fig. 4. The FM-SW mode is quite robust on the changes of the thickness of the magnetostrictive multilayer whereas the PS-SW modes are shifted to lower frequenciesdue to weaker confinement in the multilayer.

In the considered geometry (BV), the magnetoelastic interaction between SWs and R-SAWs (or S-SAWs) is exceedingly weak [8], [12], [28]. This interaction becomes noticeable for the FM-SW mode and Love SAW [18]. However, the Love SAW is hardly detectable. In our investigations, the Love SAW becomes visible only due to its hybridization with the FM-SW mode, as depicted in Fig. 4a. In the numerical simulations, we employed a simplified model that accounted only for the displacements characteristic of R-SAWs and S-SAWs, while excluding magnetoelastic coupling. Referring to previous research [12], [28], we attributed the anticrossing observed in the FM-SW mode to the interaction with the Love SAW. We demonstrated that this interaction can be tailored by adjusting the thickness of the magnetostrictive layer, thereby modifying the slope of the dispersion branches for SAWs, while having a negligible impact on the FM-SW mode. The interaction between SAWs and SWs was possible to observe because we were able to design the magnetostrictive multilayer of reduced saturation magnetization (due to the presence of non-magnetic component – Au) and with an induced static out-of-plane field (produced by cobalt layers), acting equivalently to perpendicular anisotropy.

The cobalt layers serve as a source of scattered fields that affect the dynamics of SWs in permalloy, like perpendicular anisotropy. This anisotropy reduces the FMR frequency for films magnetized in the plane, complementing the effective magnetization reduction. The cobalt subsystem exhibits a weak response to excitation by SWs in the frequency range where SAWs are observed. It can be used to adjust the SW spectra for observing interactions with SAWs. The number of layers of Co, Py, and Au in a multilayer does not affect the frequency of FMR, but it does affect the frequencies of SAWs.

The variation in the number of repetitions of the multilayer system is significant due to the localization of SAW modes within the investigated layers, thereby enabling phonon-magnon interactions. However, in the case of acoustic waves, the influence of the symmetry of the multilayer system must also be considered. The silicon substrate onto which the buffer layer and subsequently the multilayer system are deposited possesses cubic symmetry, like both the buffer layer and the multilayer system. This symmetry can be illustrated by plotting the 3D image of Young's modulus – see Fig. 5.

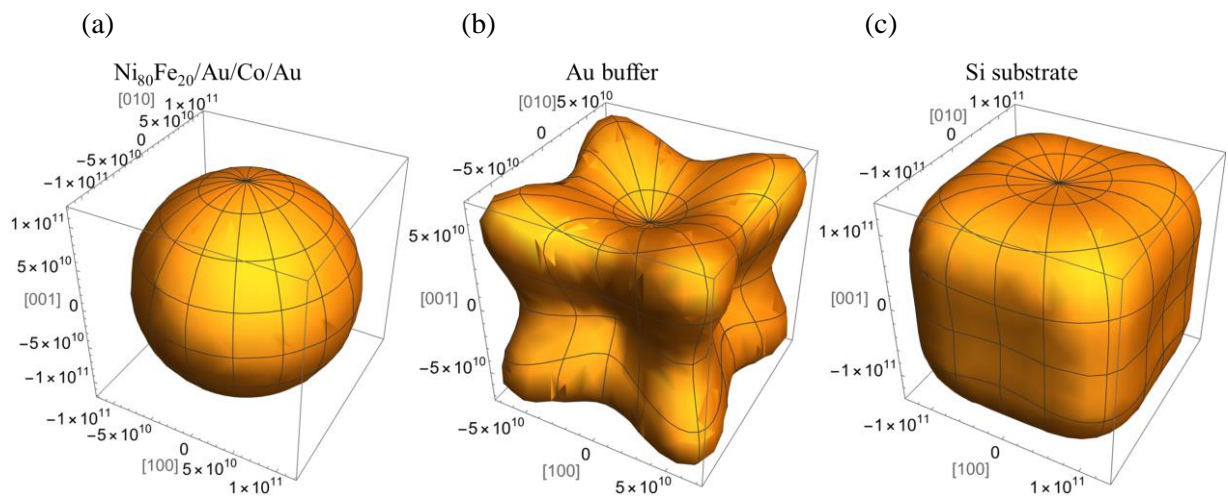

Fig. 5. The 3D visualization presents the calculated Young's modulus of the individual components within the studied system: (a) the Young's modulus of the multilayers treated as independent materials, (b) the Young's modulus of the gold buffer layer, and (c) the Young's modulus of the silicon substrate.

To calculate Young's modulus the ELATE [29] software, which is an open-source online tool for the analysis of elastic tensors, was used. For the multilayers, buffer layer and silicon substrate the elastic parameter given in the section FEM has been used. As shown, both the silicon substrate and the buffer layer exhibit strong cubic symmetry (as shown in Fig. 5). Such symmetry significantly influences the velocity of SAW propagation as well as their frequency in a specific crystallographic direction. To demonstrate the impact of the number of multilayers on the symmetry of the system, the Young's modulus was calculated for the studied materials as a function of the number of multilayer repetitions (Fig. 6). In the initial stage, we compute the elastic properties of the effective layers for different thicknesses of the $[Ni_{80}Fe_{20}/Au/Co/Au]_N$ layer. This process involves employing the proportion method, where the elastic properties of each layer are multiplied by the volume fraction of the effective layer, summed together, and then averaged across the total volume [30], [31], [32]. Then, we calculated their weighted average based on the thickness of each layer.

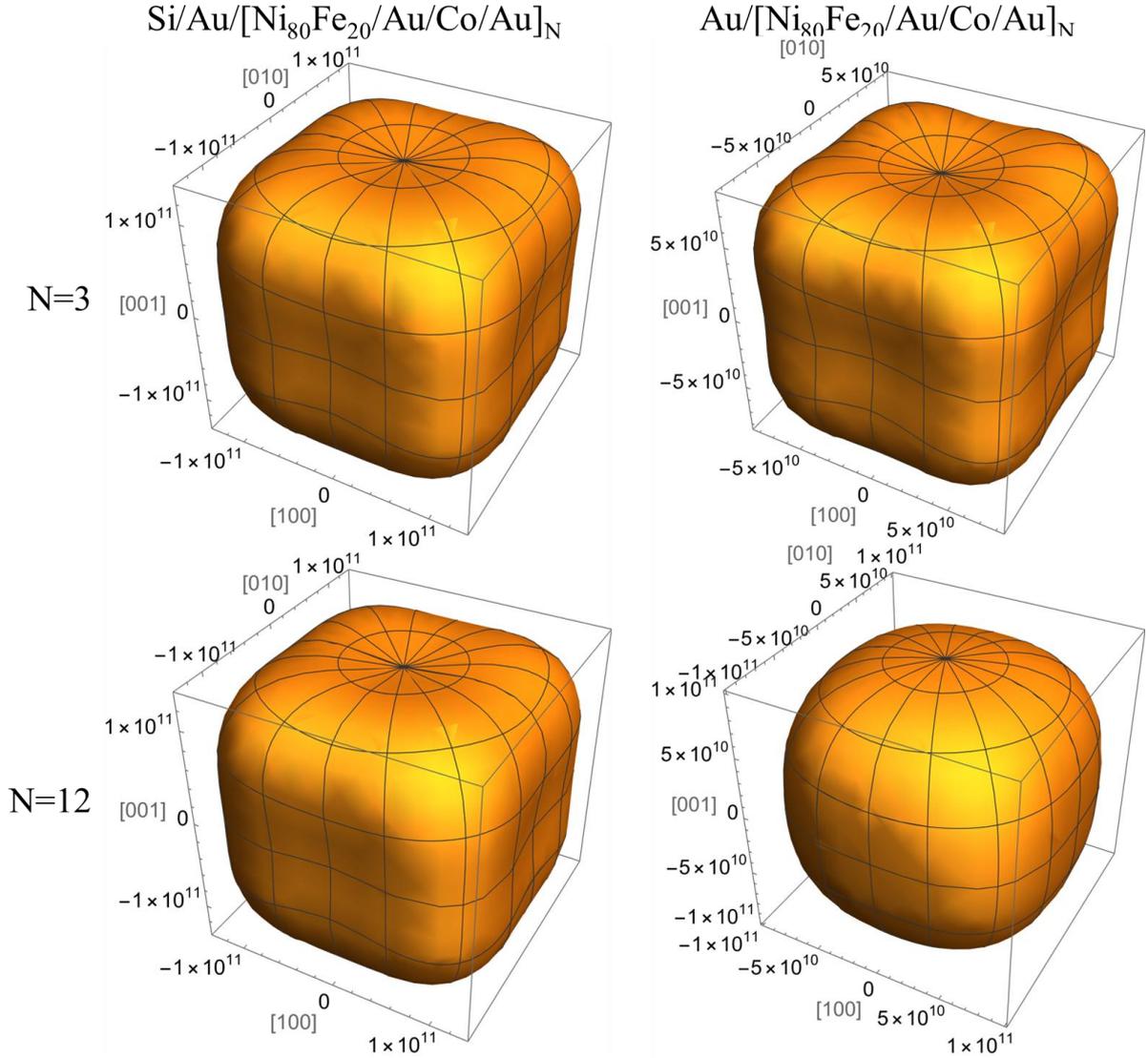

Fig. 6. The 3D visualization presents the calculated Young's modulus of the studied system: Si/Au/[Ni$_{80}$Fe$_{20}$/Au/Co/Au]$_N$ and Au/[Ni$_{80}$Fe$_{20}$/Au/Co/Au]$_N$ for $N$=3 and $N$=12 repetition.

As shown in Figure 6, the number of repetitions significantly influences the symmetry of the investigated system. This is particularly crucial for surface modes strongly localized within the multilayer structure – thus, for SAWs with large wavelengths. The multilayer structure and buffer were deposited on the (100) silicon surface. To better understand the potential effect of the anticrossing of dispersion branches of SAWs and SWs, the anisotropy of frequencies for both types of waves was determined – as shown in Fig. 7.

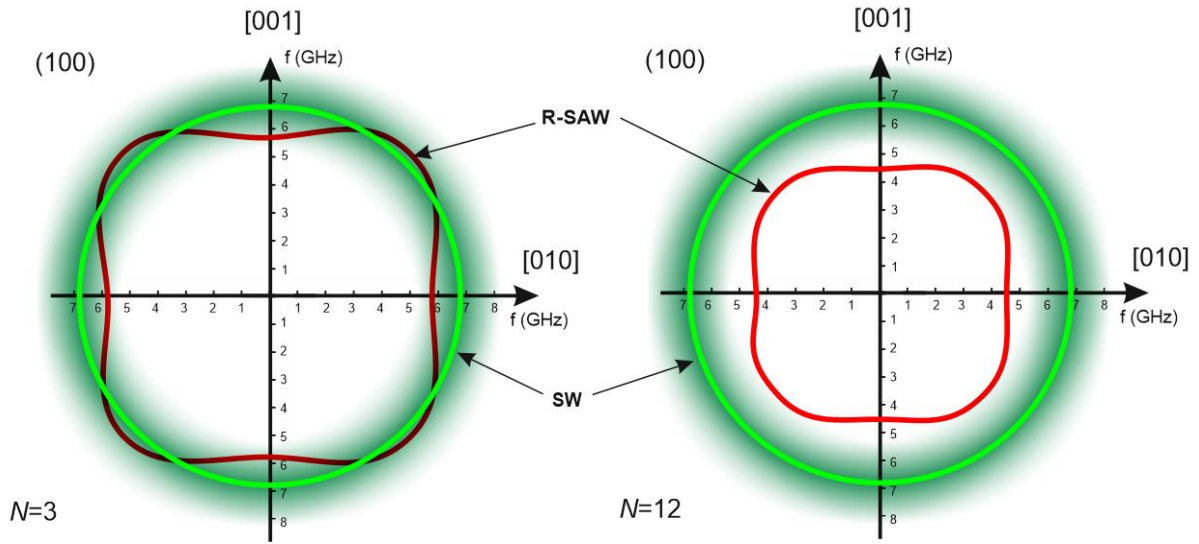

Fig. 7. The visualization of the anticrossing between R-SAW (red line) and SW (green line) for Au(30 nm)/[ $Ni_{80}Fe_{20}$/Au/Co/Au]$_N$ with $q= 0.01692$ nm$^{-1}$, under a magnetic field in BV geometry of 50 mT in the frequency ($f$) domain, is shown. The values in the parentheses represent the thickness of the Si and Au buffer layers. The crystallographic directions are depicted in the figure.

As shown in Fig. 7, the possibility of anticrossing in the system with N=3 repetitions is quite achievable under the examined measurement conditions. However, it's important to note that increasing the number of repetitions in the investigated wave vector did not result in the anticrossing of R-SAW and SW. The anisotropy of Rayleigh SAW (R-SAW) in the studied systems exhibits cubic symmetry, as depicted by the red lines in Fig. 7. The number of repetitions, denoted as N, significantly influences both the anisotropy - the anisotropy coefficient decreases by 20 percent, and the frequency of observed R-SAWs. In contrast to R-SAWs, SWs in BV geometry propagating on the (100) plane of the studied multilayer systems behave isotropically. However, the frequency of SWs is susceptible to magnetic field - potential changes in the SW frequency in Fig. 7 are illustrated as green shading and correspond to changes in the magnetic field of ±25 mT. The property of R-SAW anisotropy and the susceptibility of SW frequency to the magnetic field provide the opportunity to tune both waves in the frequency domain, thus achieving the anticrossing effect of both waves depending on the direction. The range of anticrossing can be simultaneously modulated by changing the number of repetitions of the multilayer system, affecting the SAW anisotropy coefficient on the investigated plane and by changing the length of the studied waves (changing the wave vector). It's crucial to consider the location of the individual waves in the system under consideration when exploring the potential applications of our research.

## 6. Conclusions

The experimental results demonstrate the interaction between spin waves and surface acoustic waves in a magnetic multilayer [$Ni_{80}Fe_{20}$/Au/Co/Au]$_N$ deposited on a silicon substrate with an Au buffer layer. The number of repetitions $N$ within the multilayer affects the phase velocity of surface acoustic waves, with minimal impact on the fundamental mode for backward volume spin waves. Thus, the interaction between both kinds of waves can be adjusted by modifying the dispersion of surface acoustic waves through the tuning of the multilayer thickness and the anisotropy effect of SAW.

We hypothesize that for eigenfrequencies below the FMR frequency for cobalt, the magnetization dynamics in the [Ni$_{80}$Fe$_{20}$/Au/Co/Au]$_N$ multilayer are concentrated in permalloy. Additionally, the perpendicularly magnetized cobalt layers provide an out-of-plane stray field, similar to perpendicular anisotropy, thereby reducing the FMR frequency for in-plane magnetized permalloy layers.

**Acknowledgement:**

This work was supported by Polish National Science Centre under grant no. 2020/39/D/ST3/02378.